\documentclass[aps,prb,onecolumn,floatfix,superscriptaddress,showpacs,showkeys]{revtex4}
\usepackage{epsfig}
\usepackage{amsmath}
\usepackage{hyperref}
\begin{document}
\def\il{I_{low}}
\def\iu{I_{up}}
\def\eeq{\end{equation}}
\def\ie{i.e.}
\def\etal{{\it et al. }}
\def\prb{Phys. Rev. {B }}
\def\pra{Phys. Rev. {A }}
\def\prl{Phys. Rev. Lett. }
\def\pla{Phys. Lett. A }
\def\pb{Physica B}
\def\ajp{Am. J. Phys. }
\def\mpl{Mod. Phys. Lett. {B }}
\def\ijmpb{Int. J. Mod. Phys. {B }}
\def\ijp{Ind. J. Phys. }
\def\ijpap{Ind. J. Pure Appl. Phys. }
\def\ibm{IBM J. Res. Dev. }
\def\pjp{Pramana J. Phys.}
\def\pt{Phys. Today}
\def\epl{Euro. Phys. Lett.}
\def\jpcm{J. Phys. Condensed Matter }
\def\epjb{Eur. Phys. J. B}

\title{Non-local pure spin current injection via quantum pumping and crossed Andreev reflection}
\author{Colin  Benjamin}\email{colin@sa.infn.it}
\author{Roberta Citro}
\affiliation{Dipartimento di Fisica "E. R. Caianiello",
Universita  degli Studi di Salerno, and Laboratorio Regionale
SuperMat, I.N.F.M. \\
Via S. Allende, I-84081 Baronissi (SA), Italy }

\date{\today}

\begin{abstract}
  A pure spin current injector is proposed based on adiabatic
pumping and crossed normal/Andreev reflection. The device consists
of a three-terminal ferromagnet-superconductor-semiconductor
system in which the injection of a pure spin current is into the
semiconductor which is coupled to the superconductor within a
coherence length away from the ferromagnet enabling the phenomena
of crossed normal /Andreev reflection to operate. Quantum pumping
is induced by adiabatically modulating two independent parameters
of the ferromagnetic lead, namely the magnetization strength and
the strength of coupling between the ferromagnet and the
superconductor. The competition between the normal/Andreev
reflection and the crossed normal/Andreev reflection, both induced
by pumping, leads to non-local injection of a pure spin current
into the semiconductor. The experimental realization of the
proposed device is also discussed.
\end{abstract}

 \maketitle

\section {Introduction}

\begin{figure*} [t]
\protect\centerline{\epsfxsize=4.0in \epsfbox{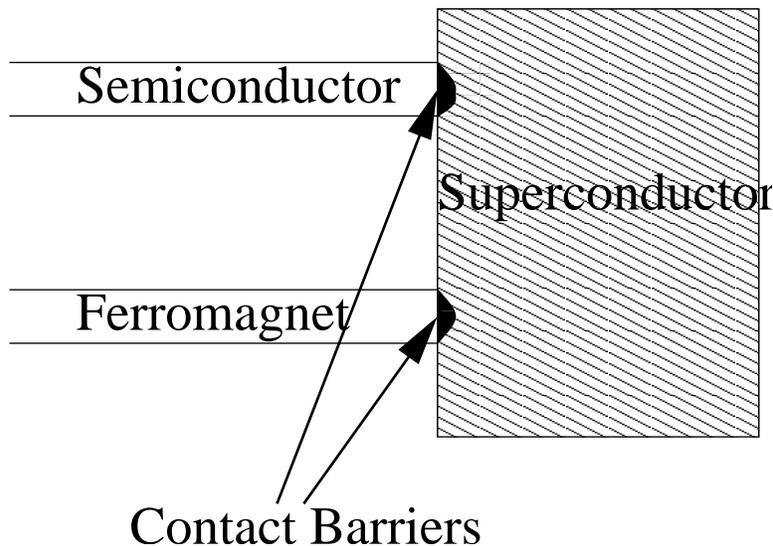}}
\caption{The proposed device: a ferromagnet and a semiconductor
kept at a small distance apart are contacted with a
superconductor. The contact barriers are indicated by the arrows.
The distance ($l$) between the ferromagnet and semiconductor leads
is much less than the coherence length ($\xi$) of the
superconductor, i.e., $l\ll \xi$.}
\end{figure*}

In the past few years there has been a lot of interest in the
field of spintronics\cite{wolf_science,prinz_science} which aims
at creating devices based on the spins of electrons. Conventional
electronics deals with charge transport and it is based on number
of charges and their energy but the performance of charge based
conventional electronics is limited in speed and dissipation. On
the contrary spintronics is based on direction of spin and spin
coupling and it is capable of much higher speed at very low
power\cite{kikkawa_nature}. Spin transport in addition is much
more resilient to impurities than charge transport since spins
wont flip until and unless there are magnetic impurities. Recent
studies have shown that spin coherence persists for hundreds of
nanoseconds over hundreds of microns\cite{sharma} and further spin
transport is largely insensitive to temperature. Further spin
based electronics promise greater integration between the logic
and storage devices and the generation, manipulation and detection
of spin currents have been the object of intense theoretical
research in recent years
\cite{aoki,tserkovnyak,chao,wang,aono,spin_wu_wang,fazio,mucciolo,chen}.
Many spintronics devices, such as the spin valve and magnetic
tunnelling junctions\cite{moodera}, are associated with the flow
of spin polarized charge currents. Spin polarized currents coexist
with charge currents and are generated when an imbalance between
spin up and spin down carriers is created, for example, by using
magnetic materials or applying a strong magnetic field or by
exploiting spin-orbit coupling in semiconductors\cite{spin-intro}.
More recently, there has been an increasing interest in the
generation of pure spin current without an accompanying charge
current\cite{sharma}. The generation of a pure spin-current is
only possible if all spin-up electrons flow in one direction and
equal amount of spin-down electrons flow in the opposite
direction. In this case the net charge current
$I_{charge}=I_{+1}+I_{-1}$ vanishes while a finite spin current
$I_{spin}=I_{+1}-I_{-1}$ exists, because $I_{+1}=-I_{-1}$, where
$I_{+1}$ or $I_{-1}$ are the electron current with spin up or spin
down.

One of the main problems with spintronics is the difficulty in
generating and transporting a spin current, i.e., spin injection
into a semiconductor\cite{spin-intro}. The ohmic injection from
ferromagnet has low efficiency because of the conductivity
mismatch and almost all of the spin polarization is lost at the
interface\cite{smidth}. Therefore pure spin currents generation
can be the most efficient tool for spin injection. One of the ways
of generating pure spin currents is through the use of quantum
pumping\cite{switkes}. In fact an experimental realization of a
quantum pumping procedure has already shown to work and generate a
pure spin current\cite{watson}. In our work a pure spin current
injection method is presented based on the principles of quantum
pumping. We propose a novel device made of a three terminal hybrid
structure, in which a ferromagnet and a semiconductor kept at a
small distance apart (less than the superconducting coherence
length), are contacted with barriers of strength $V_1$ and $V_2$
to a superconductor (see Fig.1). The quantum pumping mechanism is
incorporated by adiabatic variation of the magnetization strength
in the ferromagnetic lead and the strength of the contact barrier
at the ferromagnet-superconductor interface. It should be noted
that there is no voltage bias applied to either the ferromagnet,
the semiconductor or the superconductor, all being kept at the
same chemical potential. Adiabatic modulation of two independent
parameters is the only mechanism by which a pumped current is
generated locally in the ferromagnetic lead and more importantly
{\it nonlocally} in the semiconducting lead. As the main result of
our study we find that at a particular value of the magnetization
strength and the contact barrier strength a pure spin current can
be generated in the semiconducting lead. This effect makes
possible the use of our device as a pure spin current injector and
offers a possible solution to spin injection problems.  The
distinguishing characteristic of our proposal is non-locality,
while previous proposals generated pure spin currents in the
semiconductor locally  our work creates the same non-locally.

The two phenomena on which the operation of the proposed device
relies, are adiabatic quantum pumping and crossed normal/Andreev
reflection. Adiabatic quantum pumping is a mean of transferring
charge and/or spin carriers without applying any voltage bias by
the cyclic variation of two device control parameters. The theory
of adiabatic quantum pumping was put forth by P. W.
Brouwer\cite{brouwer}. In 1999, an adiabatic quantum electron pump
was reported in an open quantum dot where the pumping signal was
produced in response to the cyclic deformation of the confining
potential\cite{switkes}. The variation of the dot's shape squeezes
electronic wavefunction in or out of the dot thus 'pumping'
electrons from one reservoir to another\cite{polianski,cremers}.
The AC voltages applied to the quantum dot in order to change the
shape result in a DC current when the reservoirs are in
equilibrium. This non-zero current is only produced if there are
at least two time varying parameters in the system as a single
parameter quantum pump does not transfer any charge. Later on the
study of adiabatic pumping phenomenon has been extended to
adiabatic spin pumping both in experiment as well as
theory\cite{spin_wu_wang,watson}. In the experiment on quantum
spin pumping one generates a pure spin current via a quantum dot
by applying an in plane magnetic field which is adiabatically
modulated to facilitate a net transfer of spin. In addition to
investigations of pumping in quantum dots, theoretical ideas have
been put forward for spin pumping in quantum
wires\cite{sharma_qsp,citro_qsp,balent_qsp},
spin-chains\cite{shindou}, semiconductor
heterostructures\cite{mucciolo}, magnetic barriers\cite{ronald},
spin-turnstile\cite{fricot}, in presence of a superconducting
lead\cite{xing,mou,tserko}, incoherent spin pumping\cite{sela} and
in carbon-nanotubes\cite{wei}.

Crossed Andreev reflection\cite{deutscher}, on the other hand,
refers to the phenomenon when a spin up electron incident at the
ferromagnet-superconductor interface with energy below the
superconducting gap is not reflected as a spin down hole in the
same ferromagnetic lead (Andreev reflection) but is reflected in
the other lead which may be ferromagnetic/normal/semiconducting.
For this to happen, the ferromagnetic lead must be placed at a
distance {\em less than coherence length} of the superconductor
from the ferromagnetic/normal/semiconducting lead, as in Fig.1.
The phenomenon of crossed Andreev reflection can of course be
maximally enhanced when both leads are ferromagnetic with opposite
spin polarizations as shown in Ref.[\onlinecite{deutscher}] where
the distance between the ferromagnetic leads (which are
half-metals with opposite spin orientations) is neglected,
implying an effective one-dimensional model. In our work we do not
take into account the separation between the leads (the
one-dimensional model) and further the leads are not half-metals,
one is a ferromagnet while the other is a semiconductor.

 The phenomena of crossed Andreev reflection
has been explored in a large variety of systems, for details see
Ref.[\onlinecite{papers_car}]. A related phenomena which can occur
but only in the presence of a voltage bias is known as electron
cotunneling\cite{falci}. In this phenomena an electron can tunnel
into the superconductor from the ferromagnet and then again tunnel
out into the other lead placed at a distance less than the
coherence length of the superconductor. Of course this effect will
be maximally enhanced when both leads are ferromagnetic with
identical spin polarizations. In the adiabatic pumping regime the
probability of electron cotunneling is almost nil. It must be
emphasized that in a recent work\cite{chen} a spin injector was
proposed based on crossed Andreev reflection and electron
cotunneling. In Ref.[\onlinecite{chen}], competition between these
two processes leads to a pure spin current injection into the
semiconductor. In our proposed device,
 electron cotunneling ceases to operate. The only means of
transport is through quantum pumping, and because of the
competition between normal/Andreev reflection and the crossed
normal/Andreev reflection at the semiconductor/superconductor
interface a pure spin current injection into the semiconducting
lead can be obtained.

The organization of the paper is as follows: In
Sec.\ref{sec:theory} we derive the pumped current in our device by
a scattering matrix approach. In particular, we derive analytical
expressions for the charge and spin currents in the lowest order
of the contact barrier strengths. In Sec. \ref{sec:results} we
present our results for the weak pumping regime as well as the
strong pumping regime. In particular we will show that a pure spin
current injection in the semiconductor can be obtained at
particular values of the pumping parameters. Finally in
Sec.\ref{sec:setup} we discuss the possible experimental
realization of the proposed device and give the conclusions in
Sec.\ref{sec:conclusions}.

\section {Theory}
\label{sec:theory} In the device shown in Fig.~1, we consider the
pumping of charge/spin carriers by adiabatic modulation of the
magnetization strength ($h_{ex}=h_{0}+h_{p}\sin(wt)$) in the
ferromagnet and of the contact barrier strength
($V_{1}=V_{0}+V_{p}\sin(wt+\phi)$) at the
ferromagnet-superconductor interface. To study the current pumped
in such a system we apply the scattering matrix
approach\cite{beenakker,wang,blaau}. To calculate the scattering
amplitudes we start by writing the wave function for an electron
with spin $\sigma$ incident at the ferromagnet-superconductor
junction which is given by:

\begin{eqnarray}
\label{eq:wf}
\Psi_{F}&=&\binom{1}{0}e^{i p_{\sigma}^{+} x} +
\binom{0}{1}
 S_{\sigma,FF}^{he} e^{i p_{\sigma}^{-} x} + \binom{1}{0}
 S_{\sigma,FF}^{ee} e^{-i p_{\sigma}^{+} x},\mbox  {in the  ferromagnet,} \nonumber\\
 \Psi_{Sm}&=&\binom{0}{1}
S_{\sigma,SmF}^{he} e^{i q x} +
\binom{1}{0} S_{\sigma,SmF}^{ee} e^{-i q x},\mbox  {in semiconductor,}\nonumber\\
\mbox {and,}  \Psi_{Sc}&=&\alpha_{\sigma}\binom{u}{v} e^{i k^{+}
x} + \beta_{\sigma}\binom{v}{u}  e^{-i k^{-} x}\mbox { in the
superconductor}.
\end{eqnarray}
where the wavevector in the ferromagnet is given by
$p_{\sigma}^{\pm}=\sqrt{\frac{2m}{\hbar}}\sqrt{E_{F}\pm \epsilon
+\sigma h_{ex}} $, in the semiconductor is given by
$q^{\pm}=\sqrt{\frac{2m}{\hbar}}\sqrt{E_{F}\pm \epsilon}$, and in
the superconductor is given by
$k^{\pm}=\sqrt{\frac{2m}{\hbar}}\sqrt{E_{F}\pm\sqrt{\epsilon^{2}-\Delta^{2}}}$,
$h_{ex}$ represents the exchange field in the ferromagnet,
$\Delta$ being the superconducting gap and $E_F$ is the Fermi
energy. In Eq.(\ref{eq:wf}), $S_{\sigma,FF}^{he}$ is the amplitude
for Andreev reflection , $S_{\sigma,SmF}^{he}$ the amplitude for
crossed Andreev reflection , $S_{\sigma,FF}^{ee}$ the amplitude
for normal reflection, and finally $S_{\sigma,SmF}^{ee}$  the
amplitude for crossed normal reflection. $u$ and $v$ are the
superconducting coherence factors. In similar fashion one can
write the wavefunction for an electron injected with spin $\sigma$
in the semiconductor. In the Andreev
approximation\cite{andreev_approx}, we take
$k^{+}=k^{-}=k_{F}=\sqrt{\frac{2mE_{F}}{\hbar^2}}$.

The scattering amplitudes for electron/hole with spin $\sigma$
injected from either leads are calculated by applying the boundary
conditions that are set-up by matching the wave-functions  at the
interface and from current conservation at
ferromagnet-superconductor and semiconductor-superconductor
interfaces:

\begin{eqnarray}
&&\Psi_{F}(x=0)+\Psi_{Sm}(x=0)=\Psi_{Sc}(x=0),\nonumber\\
&&\frac{d\Psi_{F}}{dx}(x=0)-\frac{d\Psi_{Sc}}{dx}(x=0)=
\frac{2mV_{1}}{\hbar^2}\Psi_{Sc}(x=0),\nonumber\\
\mbox
{and,}&&\frac{d\Psi_{Sm}}{dx}(x=0)-\frac{d\Psi_{Sc}}{dx}(x=0)=
\frac{2mV_{2}}{\hbar^2}\Psi_{Sc}(x=0).
\end{eqnarray}

Solving the system of equations arising from the above boundary
conditions, all the scattering amplitudes are obtained. Similarly,
for the injection of a hole from the ferromagnetic lead, one can
determine the following scattering amplitudes $S_{\sigma,FF}^{eh}$
(for Andreev reflection), $S_{\sigma,FF}^{hh}$ (for normal
reflection), $S_{\sigma,SmF}^{eh}$ (for crossed Andreev
reflection) and $S_{\sigma,SmF}^{hh}$ (for crossed normal
reflection) in the semiconductor. The explicit expression for the
scattering amplitudes of the crossed/normal Andreev reflection can
be found in Ref.[\onlinecite{scatt}]. In the same manner one can
derive the scattering amplitudes for electron or hole injection in
the semiconducting lead. In the following we take the Fermi
energies in the semiconductor and ferromagnet to be in line with
the superconductor, hence $\epsilon=0$.

Due to the cyclic variation of external parameters $X_1$ and
$X_2$, the adiabatic pumped electron current with arbitrary spin
$\sigma$ in the semiconductor is given by:

\begin{equation}
I^{e}_{\sigma,Sm}=\frac{w q_{e}}{\pi} \int_{0}^{\tau} d \tau
[\frac{dN_{\sigma, Sm}^{e}}{dX_1}\frac{dX_1}{dt}
+\frac{dN_{\sigma, Sm}^{e}}{dX_2}\frac{dX_2}{dt}]
\end{equation}
wherein, $\tau=2\pi/w$ is the cyclic period, $w$ is the pumping
frequency and $q_e$ represents the electronic charge.
$\frac{dN_{\sigma, Sm}^{e}}{dX_j}$ is the electronic
injectivity\cite{buttiker,wang} in the semiconducting lead given
at zero temperature by

\begin{equation}
\frac{dN_{\sigma, Sm}^{e}}{dX_j}=\frac{1}{2\pi} \sum_{\beta=F,Sm}
\Im[S_{Sm\beta,\sigma}^{*,ee}\partial_{X_j}S_{Sm\beta,\sigma}^{ee}
+S_{Sm\beta,\sigma}^{*,eh}\partial_{X_j}S_{Sm\beta,\sigma}^{eh}]
,\mbox { with, j=1,2}
\end{equation}
where, $X_{1}=V_{1}$, and $X_{2}=h_{ex}$. $\Im$ denotes the
imaginary part of the quantity in parenthesis. The first term is
the injectivity of the electron due to the variation of the
modulated parameter, i.e. the partial density of states (DOS) for
an electron coming from either the ferromagnet or semiconductor
and exiting the system as an electron in the semiconductor, and
the second term is the injectivity of a hole, i.e., the DOS for a
hole coming from the semiconductor or ferromagnet and exiting the
system as an electron in the semiconductor. Similarly, one can
calculate the pumped hole current in the semiconducting lead and
it is given by the expression, with $q_e$ replaced by the hole
charge $q_h$ and with $e$ replaced by $h$ in Eq.4, as below:

\begin{equation}
I^{h}_{\sigma,Sm}=\frac{w q_{h}}{\pi} \int_{0}^{\tau} d \tau
[\frac{dN_{\sigma, Sm}^{h}}{dX_1}\frac{dX_1}{dt}
+\frac{dN_{\sigma, Sm}^{h}}{dX_2}\frac{dX_2}{dt}].
\end{equation}
$\frac{dN_{\sigma, Sm}^{h}}{dX_j}$ is the hole injectivity in the
semiconducting lead given at zero temperature by

\begin{equation}
\frac{dN_{\sigma, Sm}^{h}}{dX_j}=\frac{1}{2\pi} \sum_{\beta=F,Sm}
\Im[S_{Sm\beta,\sigma}^{*,hh}\partial_{X_j}S_{Sm\beta,\sigma}^{hh}
+S_{Sm\beta,\sigma}^{*,he}\partial_{X_j}S_{Sm\beta,\sigma}^{he}]
\end{equation}

In the weak pumping regime i.e., $h_{p} \ll h_{0}$ and $V_{p}\ll
V_{0}$, the adiabatically pumped electron current into the
semiconducting lead can be written as below\cite{blaau}, with
$z_{i}=\frac{2mV_i}{k_F\hbar^2}(i=0,1,2,p)$, $h^{'}=h_{p}/E_{F}$
and $h=h_{0}/E_{F}$:

\begin{equation}
I_{\sigma,Sm}^{e}=\frac{wq_{e}\sin(\phi)z_{p}h^{'}}{\pi}
\sum_{\beta=F,Sm}\Im[\partial_{z_0}S_{\sigma,Sm\beta}^{*,ee}\partial_{h}S_{\sigma,Sm\beta}^{*,ee}
+\partial_{z_0}S_{\sigma,Sm\beta}^{*,eh}\partial_{h}S_{\sigma,Sm\beta}^{*,eh}].
\end{equation}

Similarly, the adiabatically pumped hole current into the
semiconducting lead can be written as:

\begin{equation}
I_{\sigma,Sm}^{h}=\frac{wq_{h}\sin(\phi)z_{p}h^{'}}{\pi}
\sum_{\beta=F,Sm}\Im[\partial_{z_0}S_{\sigma,Sm\beta}^{*,hh}\partial_{h}S_{\sigma,Sm\beta}^{*,hh}
+\partial_{z_0}S_{\sigma,Sm\beta}^{*,he}\partial_{h}S_{\sigma,Sm\beta}^{*,he}].
\end{equation}

In the following, $q_{e}=-q_{h}$. Further, since for s-wave
superconductor the scattering amplitudes satisfy the condition
$(S_{\sigma,SmSm}^{hh})^{*}=S_{\sigma,SmSm}^{ee},
(S_{\sigma,SmF}^{hh})^{*}=S_{\sigma,SmF}^{ee}$, and
$(S_{\sigma,SmSm}^{eh})^{*}=-S_{\sigma,SmSm}^{he},
(S_{\sigma,SmF}^{eh})^{*}=-S_{\sigma,SmF}^{he}$ (where $*$ denotes
complex conjugation of the scattering amplitude), the pumped hole
and electron current are exactly the same, and because of this we
only derive the expressions for the pumped electron current with
arbitrary spin index understanding that it is exactly equal to the
pumped hole current. Similar to the above expressions one can
write the pumped electron/hole current with arbitrary spin index
into the ferromagnetic lead. Since our focus is the use of this
device as a spin injector, we confine ourselves to the
semiconducting lead only.

From the reflection amplitudes (crossed
Andreev/normal)\cite{scatt} and in the weak pumping regime (see
Eq. 7), one can derive the pumped electron current, charge and
spin currents in the semiconducting lead up-to first order in
contact barrier strengths $z_0=\frac{2mV_0}{k_F\hbar^2}$ and
$z_2=\frac{2mV_2}{k_F\hbar^2}$ (assuming $z_0$ and $z_2 $ are
small). The explicit expression for the current is (with ,
$h=h_{0}/E_{F}$):
\begin{equation}
\label{eq:currsigma}
I_{\sigma}=I_{\sigma,Sm}^{e}/I_{0}=\frac{-1}{1-h^2}[A+\sigma
B_{\sigma}+C_{\sigma}]/D^{2},
\end{equation}
where
\begin{eqnarray}
I_{0}&=&(wq_{e}\sin(\phi)z_{p}h')/\pi,\nonumber\\
D&=&-7-2(\sqrt{1-\sigma h}+\sqrt{1+\sigma
h})(1+2\sqrt{1-h^{2}})+6\sqrt{1-h^{2}}-4h^{2} \nonumber \\
A&=&16h(1-h^{2})+14h \sqrt{1-h^{2}},\nonumber\\
B_{\sigma}&=&\sqrt{1-h^2}(2+4h^{2})-14(1-h^{2})-\sqrt{1+\sigma
h}(10+12h^{2})\sqrt{1-\sigma h}(-4+8h^{2})\nonumber\\ &+&
\sqrt{1+\sigma h}\sqrt{1-h^{2}}(-4-48h^{2})+\sqrt{1-\sigma
h}\sqrt{1-h^{2}}(5-36h^{2}),\nonumber\\
C_{\sigma}&=&\sqrt{1-\sigma h}(36h^{3}-32h)+\sqrt{1+\sigma h}
(-36h^{3}+58h)-\sqrt{1-\sigma h}\sqrt{1-h^{2}}(46h)\nonumber\\
\label{eq:coeff} &+&\sqrt{1+\sigma h}\sqrt{1-h^{2}}(60h).
\end{eqnarray}
One should notice that up to first order there are no terms
involving $z_0$ and $z_2$  as the first terms involving $z_0$ or
$z_2$ that appear in the expansion are of order $O(z_i^2), i=0,2$.
From Eqs. (\ref{eq:currsigma}-\ref{eq:coeff}) the charge and spin
current pumped into the semiconducting lead are derived as below:
\begin{eqnarray}
I_{spin}=I_{+1}-I_{-1}&=&\frac{-1}{1-h^{2}}[-28(1-h^{2})+X(-4h^{2}-14+Y(1-84h^{2}))-W(26h-106hY)]/D^{2},\\
I_{charge}=I_{+1}+I_{-1}&=&\frac{-1}{1-h^{2}}[
28hY+32h(1-h^{2})-W(6+20h^{2}+(9+12h^{2})Y)-X(26h+14hY)]/D^{2}.
\end{eqnarray}
where, $X=\sqrt{1+h}+\sqrt{1-h}, Y=\sqrt{1-h^{2}},
W=\sqrt{1+h}-\sqrt{1-h}$ and $D$ as in Eq. (10). At the value of
the exchange field $h=0$, the charge current is zero while the
spin current is finite. Let us note that since we considered a
one-dimensional model (see also Ref.[\onlinecite{deutscher}]) the
distance $l$ between the ferromagnet and the semiconductor does
not appear explicitly in the result for the pumped current,
further the width of the ferromagnetic and semiconducting leads is
not taken into account. In Ref.[\onlinecite{yamashita}] both these
characteristics have been incorporated and it has been explicitly
shown that crossed Andreev reflection ceases to be effective in
the limit when the separation between the leads approaches the
superconducting coherence length. Although the considered
one-dimensional transport model we have considered is simplistic,
it captures the main physics of crossed Andreev reflection.

\begin{figure*} [!]
\protect\centerline{\epsfxsize=7.0in \epsfbox{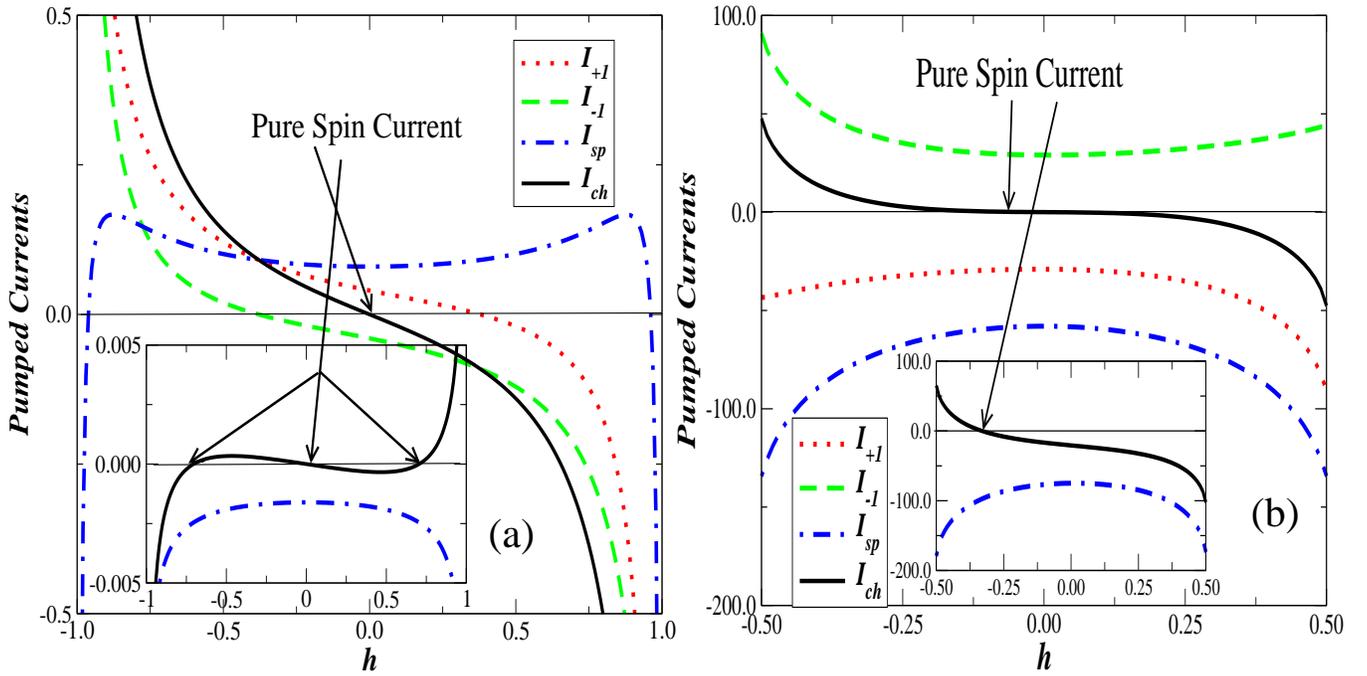}}
\caption{(Color online) The pumped charge and spin currents into
the semiconducting lead as function of the  dimensionless
magnetization strength $h=h_{0}/E_{F}$ in the ferromagnet. (a) The
weak pumping regime. Parameters are $z_{0}=z_{2}=0$. In the inset
parameters are $z_{0}=2, z_{2}=0$. (b) The strong pumping regime.
Parameters are $z_{0}=z_{2}=0, z_{p}=4.0$, $h^{'}=0.5$ and
$\phi=\pi/2$. In the inset parameters are $z_{1}=2, z_{2}=0,
z_{p}=4.0, h^{'}=0.5$ and $\phi=\pi/2$. Herein $h^{'}=h_{p}/E_{F}$
and $z_{i}=2mV_{i}/\hbar^{2}k_{F}$ with $i=0,1,2,p$. The arrows
indicate the specific places wherein pure spin current flows in
the semiconductor. $I_{\pm}$ denotes the spin up and down
currents.}
\end{figure*}

\begin{figure*} [!]
\protect\centerline{\epsfxsize=7.0in \epsfbox{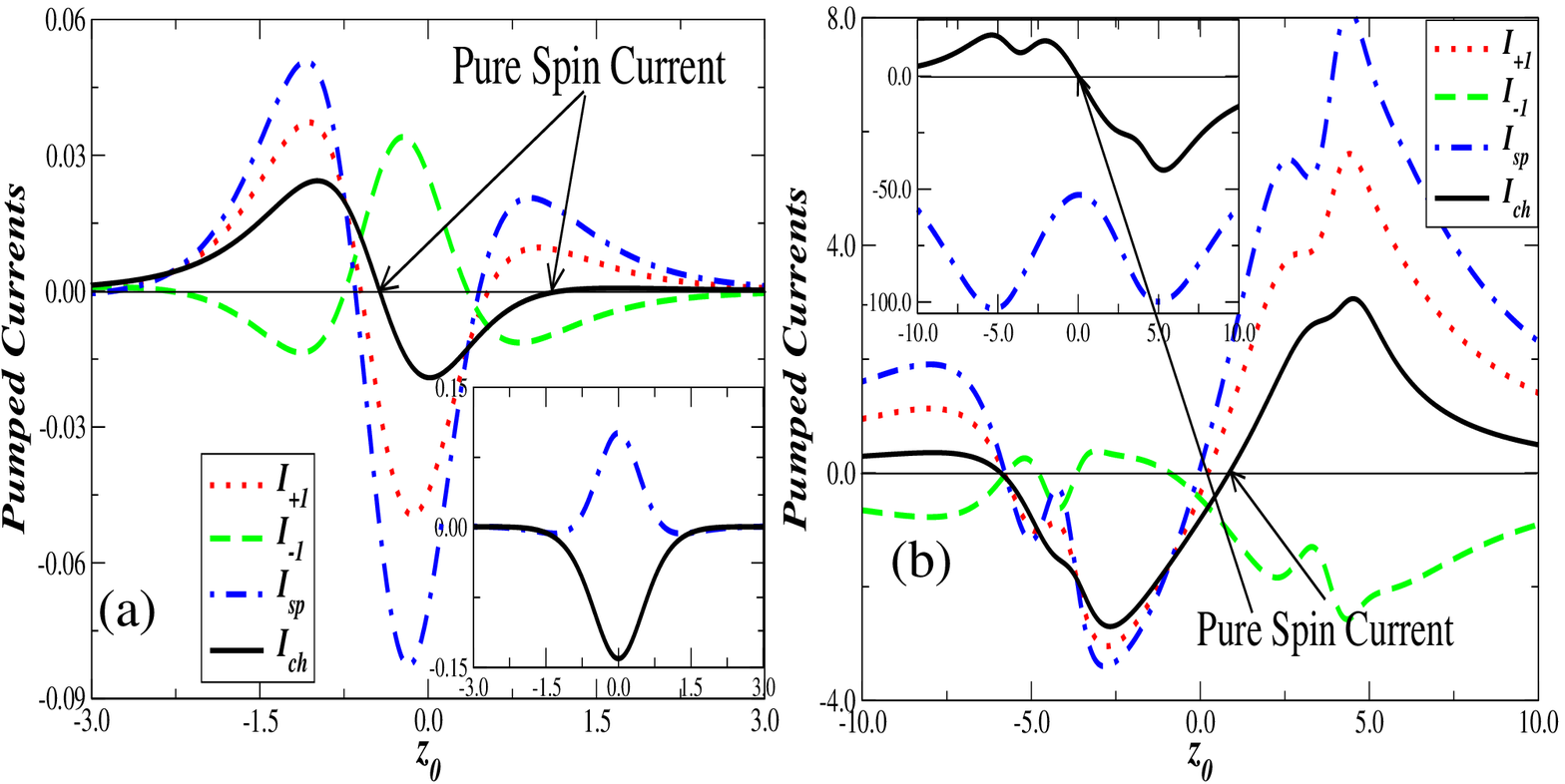}}
\caption{(Color online) The pumped charge and spin currents into
the semiconducting lead as function of the contact barrier
strength $z_{0}$. (a) The weak pumping regime. Parameters are
$z_{2}=0.0, h=0.5$. In the inset parameters are $z_{2}=2.0,
h=0.5$. (b) The strong pumping regime. Parameters are $z_{2}=2.0,
z_{p}=4.0$, $h^{'}=0.45, h=0.1$ and $\phi=\pi/2$. In the inset
parameters are $z_{2}=0.0, z_{p}=4.0, h^{'}=0.45, h=0.1$ and
$\phi=\pi/2$. The arrows indicate the specific places wherein pure
spin current flows in the semiconductor.}
\end{figure*}

\begin{figure} [!]
\protect\centerline{\epsfxsize=7.0in \epsfbox{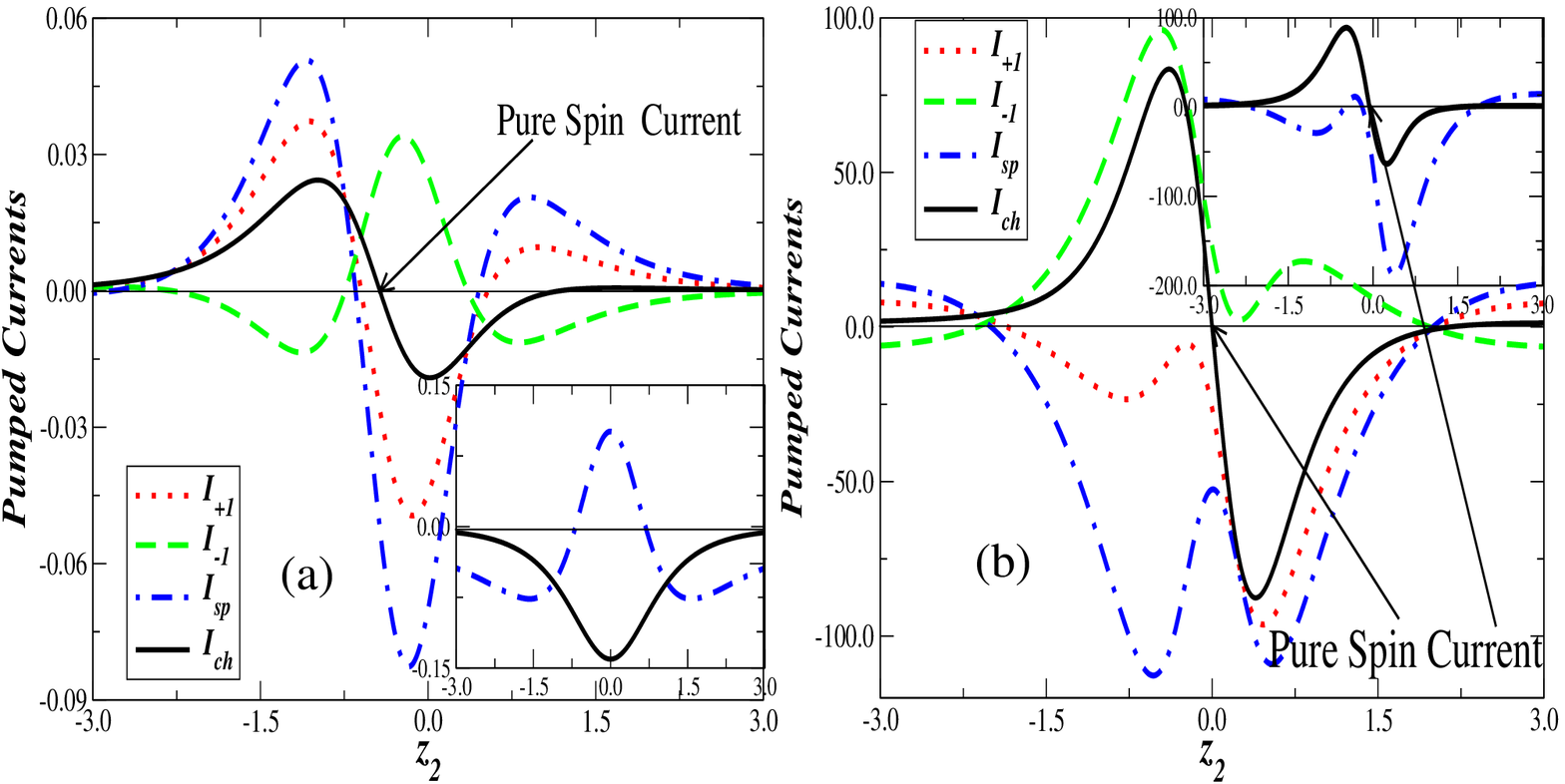}}
\caption{(Color online) The pumped charge and spin currents into
the semiconducting lead as function of the contact barrier
strength $z_{2}$. (a) The weak pumping regime. Parameters are
$z_{0}=2.0, h=0.5$. In the inset parameters are $z_{0}=0.0,
h=0.5$. (b) The strong pumping regime. Parameters are $z_{0}=2.0,
z_{p}=4.0$, $h^{'}=0.45, h=0.1$ and $\phi=\pi/2$. In the inset
parameters are $z_{0}=0.0, h=0.1, z_{p}=4.0, h^{'}=0.45$ and
$\phi=\pi/2$. The arrows indicate the specific places wherein pure
spin current flows in the semiconductor. }
\end{figure}
\section{Characteristics of the pumped charge and spin currents}
\label{sec:results}

The results for the pumped charge and spin currents both in the
weak as well as strong pumping  regimes are plotted in the Figs.
2-6. In all the figures the parameters are in their dimensionless
form. The weak pumping regime can be described through Eqs.(9-12)
and it is shown in Figures 2(a), 3(a) and 4(a), as well as through
Eq.~(4) as shown in Figure 6(a). In Figures 2(a), 3(a) and 4(a)
the currents are scaled by a factor $I_{0}$, while in Figures
2(b), 3(b), 4(b) and in Figures 5 and 6, they are in units of
$wq_{e}/\pi$ and are multiplied by (-100) for better visibility.

In Fig.2(a), we plot the exact results in the regime where contact
barrier strengths are neglected [see Eqs. 9-12]. The figure shows
that at the value of the normalized magnetization strength
$h=0.0$, we have zero charge current and a finite spin current.
The arrows denote the points where the charge current is zero. In
the inset of Fig. 2(a) the pumped spin and charge currents are
plotted by taking the contact barrier strength at the
ferromagnet-superconductor into account while neglecting the
contact barrier strength at the semiconductor-superconductor
junction. The pumped charge current is zero for three distinct
values of the magnetization strength. In Fig.2(b), the results for
the strong pumping regime are reported. As before the contact
barrier strengths are neglected in the main figure, while the
results in presence of the contact barrier at the
ferromagnet-superconductor junction are shown in the inset. Not
much difference between the weak and strong pumping regimes is
observed: in both cases at particular values of the magnetization
strength a pure spin current is seen. In Fig.~3(a) we plot the
pumped currents along with the pumped spin and charge currents as
function of the normalized contact barrier strength $z_0$ at the
ferromagnet-superconductor interface. The inset shows the currents
when the contact barrier strength at the
semiconductor-superconductor junction is taken into account.
Herein, at particular values of $z_0$ a pure spin current in the
semiconductor is observed, while in the inset there is no pure
spin current. In Fig.~3(b), the currents for the strong pumping
case are plotted. For this case whether or not the contact barrier
strength is taken into account, there is a pure spin current.
 In Fig.~4(a), we plot the
pumped currents along with the pumped spin and charge currents as
function of the normalized contact barrier strength $z_2$ at the
semiconductor-superconductor interface. The inset shows the
currents when the contact barrier strength at the
ferromagnet-superconductor junction is neglected. Herein, at a
particular value of $z_2$ a pure spin current in the semiconductor
is observed, but in the inset no such value occurs. In Fig.~4(b),
the currents for the strong pumping case are plotted. In both
Fig.~4(b) as well as its inset a pure spin current occurs.

\begin{figure} [!]
\protect\centerline{\epsfxsize=7.0in \epsfbox{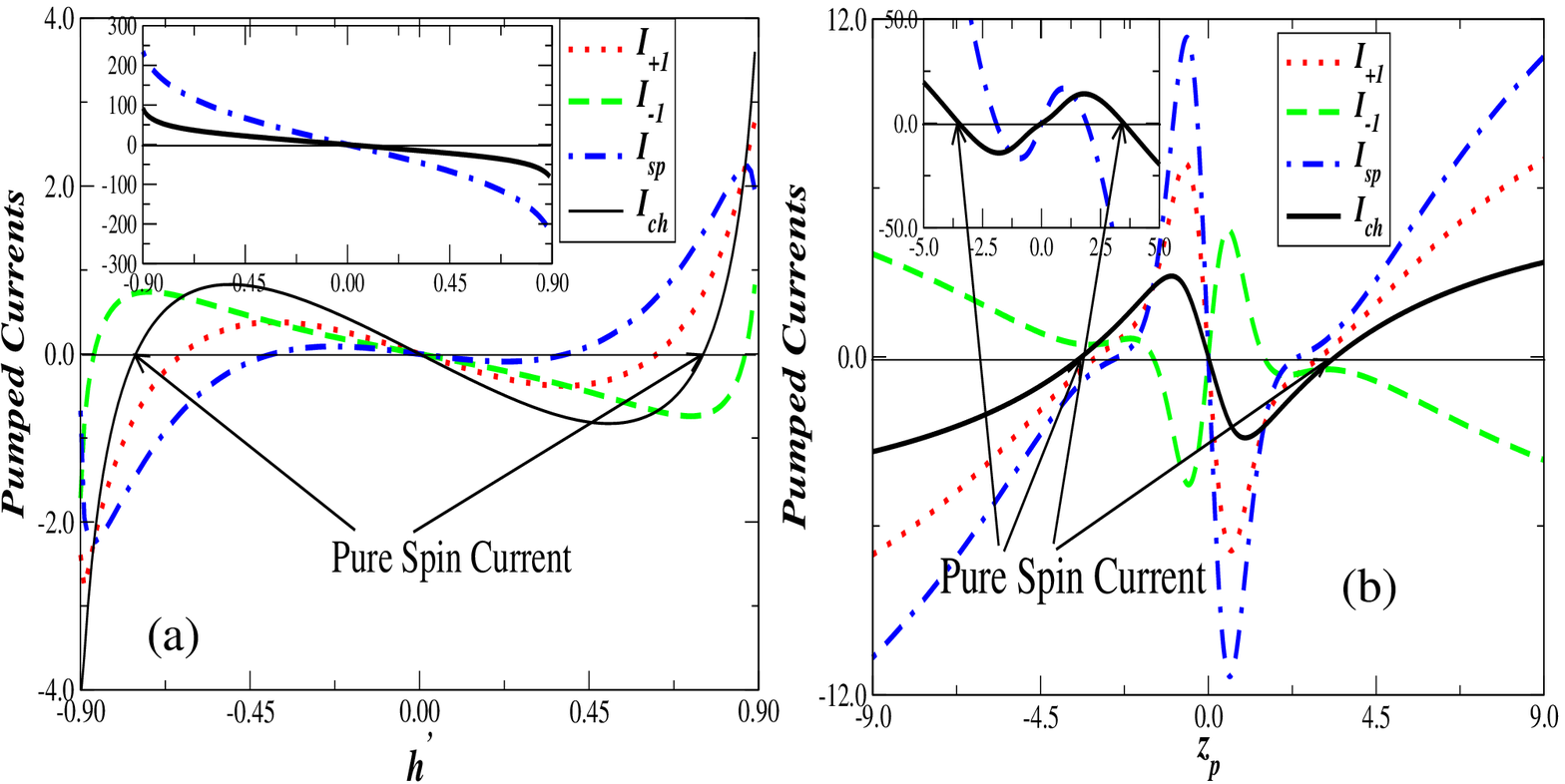}}
\caption{(Color online) The pumped currents as function of the
strength of pumping parameters. (a) The pumped currents as
function of the magnetization amplitude. Parameters are
$z_{1}=0.0, z_{2}=2.0, z_{p}=4.0, h=0.1$ and $\phi=\pi/2$. In the
inset parameters are $z_{1}=2.0, z_{2}=0.0, z_{p}=4.0, h=0.1$ and
$\phi=\pi/2$.(b)The pumped currents as function of the amplitude
of pumped contact barrier strength at the
ferromagnet-superconductor junction. Parameters are $z_{0}=0.0,
z_{2}=2.0, h_{p}=.8, h=0.1$ and $\phi=\pi/2$. In the inset
parameters are $z_{0}=0.0, z_{2}=0.0, h_{p}=.8, h=0.1$ and
$\phi=\pi/2$.  The arrows indicate the specific places in the
parameter regime wherein pure spin current flows in the
semiconductor.}
\end{figure}

\begin{figure} [!]
\protect\centerline{\epsfxsize=7.0in \epsfbox{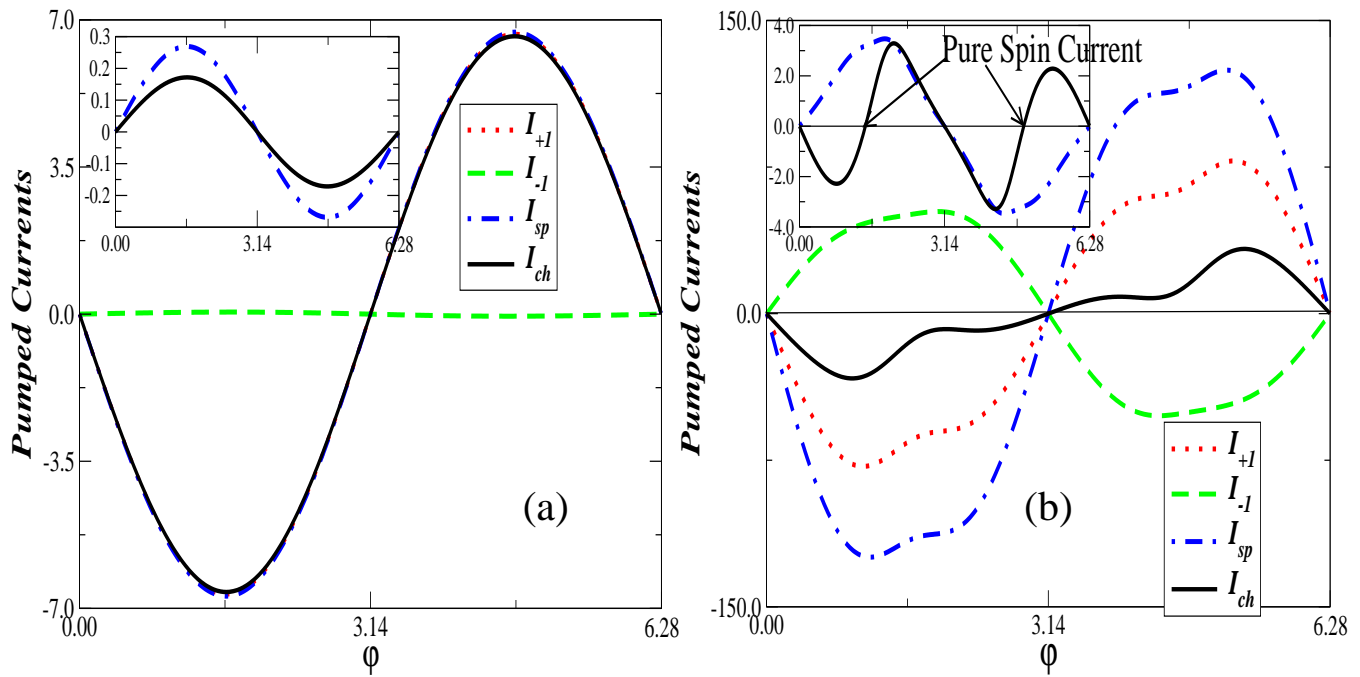}}
\caption{(Color online) The pumped currents as function of the
phase difference. (a) for weak pumping, the parameters are $h=0.8,
z_{1}=4.0, z_{2}=0.0, h^{\prime}=0.1$ and $z_{p}=0.4$. Notice that
the down spin current is almost zero and the pumped currents are
almost sinusoidal. In the inset, parameters are $h=0.8, z_{1}=4.0,
z_{2}=2.0, h^{\prime}=0.1$ and $z_{p}=0.4$. (b) For very strong
pumping, the parameters are $h=0.1, z_{0}=.4, z_{2}=0.0,
h^{\prime}=0.8$ and $z_{p}=4.0$. In the inset the parameters are
$h=0.1, z_{0}=.4, z_{2}=2.0, h^{\prime}=0.8$ and $z_{p}=0.4$.
Notice that the sinusoidal dependence of the pumped currents is
lost.}
\end{figure}

The dependence of the pumped currents on the strength of modulated
parameter is another crucial indicator of the regime parameters in
which the pump operates as a pure spin current injector. With this
in mind in Fig.~5, we plot the pumped currents and the spin and
charge currents as function of the strength of the modulated
parameter, in (a) the magnetization and in (b) the contact barrier
strength. A pure spin current appears at particular values of the
modulated magnetization in Fig.5(a) and at particular values of
the modulated contact barrier strength in Fig.5(b).

 In Fig.~6, we plot the pumped spin and charge currents as a
function of the phase difference $\phi$. In Fig.~6(a), the weak
pumping regime is shown. In this regime the pumped currents are
almost sinusoidal. For the values of the parameters considered in
the figure, the down spin current is almost zero and an almost
pure up spin current is obtained. In Fig.~6(b), the strong pumping
regime is shown. As expected, the sinusoidal dependence on the
phase is lost. Furthermore in the inset of Fig.6(b) we see that
for the parameters taken into account one has pure spin current at
$\phi=\pi/2$ and $3\pi/2$.  This is quite an important result
since in this case both the contact barrier strengths are not
neglected. A question that could arise is related to the magnitude
of the pumped spin current. From the previous results one clearly
sees that the pumped currents are noticeably larger in the strong
pumping regime. The device thus should be ideally suitable for use
in the strong pumping regime. For the parameters considered in
Figure 6, taking the frequency $w$ around $100 MHz$ as in
Ref.[\onlinecite{switkes}], the order of magnitude estimate of the
pumped current is around $10^{-11}$ Amperes, a value detectable in
present day experiments. To summarize the results from all the
figures and to make sense of all the parameters used, we in Table
I list the physical parameters , their values (range as used in
the figures), their physical meaning and materials wherein these
can be realized.

\begin{table*}[h]
\begin{center}
\begin{tabular}{|c|c|c|c|c|c|}
\hline
   Parameter$\downarrow$& Symbol &Figure& Values& Physical
meaning&  materials    \\ \hline Contact barrier strength&
$z_{i}$ & 3 and 4 &$0$ & ballistic contact& sharvin contacts  \\
\hline Contact barrier strength & $z_{i}$ & 3 and 4 &$\neq 0 $&
non reflectionless contacts & tunnel contacts
\\\hline
Magnetization & h & 2 &$\sim E_{F}/2$ & $50\%$ polarized & NiFe,
Ni, Co\\ \hline Superconducting coherence length& $\xi$& 1 &$\xi
\gg l $ & crossed andreev reflection & s-wave superconductors
\\
& & & & & (eg. Aluminium in Ref.[\onlinecite{beckmann}].)\\\hline
\end{tabular}
\caption{\small{  A comparative analysis of parameters,values and
materials}}\vspace{-9mm}
\end{center}
\end{table*}
\vskip 0.1in

 In the table, above a sharvin contact is defined when
strength of contact barrier is equal to zero. These type of
contacts can be realized when a point contact has a size $d$
smaller than mean free path $l$. This implies completely ballistic
transport through the contact. In this case an electron is either
Andreev reflected in the ferromagnet or cross Andreev reflected
into the semiconductor. For non-zero contact barrier strength, the
contacts are defined as tunnel contacts or non reflectionless
contacts, and in this case in addition to Andreev reflection in
ferromagnet and cross Andreev reflection in semiconductor there
can be normal reflection in ferromagnet and crossed normal
reflection in semiconductor. The finite magnetization $h$ in the
proposed device can be obtained by using any type of ferromagnetic
material e.g., NiFe, Cobalt or Nickel. Lastly the superconductor
would ideally be of s-wave type since it has large coherence
length. A high T$_{c}$ superconductor could also be used in the
device but in that case the distance between the leads would have
to be very small.

\section{Experimental Realization}
\label{sec:setup}

The experimental realization of the proposed device is not
difficult. The phenomenon of crossed Andreev reflection has been
demonstrated in two recent experiments\cite{beckmann,russo}. In
the experiment of Ref.[\onlinecite{beckmann}] a sample geometry
consisting of an aluminum bar with two or more ferromagnetic wires
forming point contacts has been considered. By measuring the
nonlocal resistance in the superconducting state of such
structures a spin-valve signal has been observed whose sign,
magnitude and decay length scale are consistent with predictions
made for crossed Andreev reflection. Our suggestion is to
integrate the quantum pumping mechanism  (notably seen in four
experiments till date, see
Refs.[\onlinecite{switkes,lorke,watson,dicarlo}]) into such type
of set-up. This can be easily done: the magnetization can be
adiabatically modulated by an external magnetic field while the
strength of the contact barrier at the ferromagnet-superconductor
interface can be modulated by applying a suitable gate voltage at
the junction. This procedure should enable a nonlocal pure spin
current generation in the semiconducting lead. The detection of
the pure spin current could be achieved through a quantum spin
Hall effect set-up\cite{spin-hall}. Apart from this method, the
spin pump can be connected to a (semiconducting) ferromagnet with
a known magnetization direction\cite{psharma}, or a
gate-controlled bidirectional spin filter \cite {folk} could also
be used to detect the spin current.

\section{Conclusions}
\label{sec:conclusions}

 A novel device for pure spin current
generation based on the interplay of adiabatic pumping and crossed
Andreev/normal reflection in a three-terminal
ferromagnet-superconductor-semiconductor system has been proposed.
The transfer of charges/spins in the device is achieved by
adiabatic quantum pumping without any voltage bias applied.
Varying the strength of the pumping parameters, namely the
magnetization strength in the ferromagnet and the contact barrier
strength at the ferromagnet-superconductor junction, we have shown
that a pure spin current can be injected into the semiconducting
lead in a completely nonlocal way. As already mentioned in the
Introduction there are many different ways to inject a spin
current into a semiconductor. One of the most commonly used
techniques is that of using ferromagnets. This technique has been
shown to be very inefficient as almost all of the spin
polarization is lost at the interface. In this work we have used
the method of quantum pumping to create a pure spin current in the
semiconductor itself. Quantum pumping methodology has been used in
many recent works to inject spin currents, however our work is
novel in two respects. Firstly, it invokes non-locality. Almost
all proposals which we are aware of, invoking quantum pumping to
induce a spin current are local, i.e., time dependent voltages are
applied directly to the  semiconductor. What our proposal proves
is that, though one does not touch the semiconductor, none the
less one generates a pure spin current in  the semiconductor.
Secondly, no ferromagnet-semiconductor interface is used and
therefore the problem of resistivity mismatch is avoided. As we
have shown, the proposed device ideally operates in the strong
pumping regime where the spin current is noticeably larger than
the charge current. The experimental realization of such device as
a pure spin current injector has also been discussed.

\section{Acknowledgments}
The authors would like to thank Prof. Costabile G. and  Carapella
G. for useful discussions.


\begin{thebibliography}{99}
\bibitem{scatt} The scattering amplitudes required to calculate the
pumped current in the semiconducting lead are given by:
\begin{eqnarray*}
S^{ee}_{\sigma,SmF}&=&\frac{-2(1+\sigma h_{1})^{1/4}(1+\sqrt{1-\sigma h_{1}}+i(z_{1}+z_{2}))}{D},S^{he}_{\sigma,SmF}=\frac{2(1+z_{2}^{2})(z_{1}-i \sqrt{1-\sigma h})}{(1+i z_{2})D},\\
S^{he}_{\sigma,SmSm}&=&\frac{2[z_{1}P+z_{2}(z_{1}^{2}+Q)+i\{z_{1}z_{2}P-z_{1}^{2}-Q\}]}{(1+iz_{2})D}\\
S^{ee}_{\sigma,SmSm}&=&\frac{[1-P-2z_{1}z_{2}(P+1)-z_{2}^{2}(1+Q+z_{1}^{2})+i\{2z_{1}+2z_{2}-Pz_{2}(1+z_{1}z_{2})+2z_{2}(Q+z_{1}^{2})\}]}{D}\\
& &\mbox {with $ Q=\sqrt{1-h_{1}^2}, P=\sqrt{1+\sigma h_{1}}-\sqrt{1-\sigma h_{1}}$,  $h_{1}=h_{ex}/E_{F}$,}\\
\mbox {and }
D&=&1+3z_{1}z_{2}+(2+z^{2}_{2})(Q+z^{2}_{1})+\sqrt{1-\sigma
h_{1}}+\sqrt{1+\sigma h_{1}}+iP(z_{1}(2+z^{2}_{2})+z_{2}).
\end{eqnarray*}
\end{thebibliography}
\end{document}